# DETECTION OF THE ARTERIAL INPUT FUNCTION USING DSC-MRI DATA

**Alkhimova Svitlana**
Associate Professor of the Department of Biomedical Cybernetics, PhD
National Technical University of Ukraine "Igor Sikorsky Kyiv Polytechnic Institute",
Kyiv, Ukraine

**Sazonova Kateryna**
Student of the 6th course
National Technical University of Ukraine "Igor Sikorsky Kyiv Polytechnic Institute",
Kyiv, Ukraine

*Abstract. Accurate detection of arterial input function is a crucial step in obtaining perfusion hemodynamic parameters using dynamic susceptibility contrast-enhanced magnetic resonance imaging. It is required as input for perfusion quantification and has a great impact on the result of the deconvolution operation. To improve the reproducibility and reliability of arterial input function detection, several semi- or fully automatic methods have been proposed. This study provides an overview of the current state of the field of arterial input function detection. Methods most commonly used for semi- and fully automatic arterial input function detection are reviewed, and their advantages and disadvantages are listed.*

**Keywords:** arterial input function; deconvolution; hemodynamic parameters; quantitative perfusion; dynamic susceptibility contrast-enhanced perfusion; magnetic resonance imaging.

**Introduction**

Dynamic susceptibility contrast (DSC) MRI is one of the most commonly used perfusion techniques, which is based on the susceptibility-induced signal loss on ultrafast $T2^*$- or T2-weighted sequences resulting from the passage of a bolus of contrast through the capillary bed [1].

DSC-MRI is widely used when examining patients with oncological diseases. It allows examining in detail tumors and measuring such tumor pathophysiology as vascular permeability, vessel caliber, tumor cell size, and cytoarchitecture [2, 3]. It is more useful than standard MRI to detect a tumor at an early stage of its formation when the chances of treating the patient are very high. In addition, DSC-MRI can be used for the assessment of ischemic stroke, neurovascular disease, and neurodegenerative disorders [3, 4].

The most commonly calculated physiological and hemodynamic-related parameters are cerebral blood flow (CBF), cerebral blood volume (CBV), mean transit time (MTT), time-to-maximum of the residue function (Tmax) and so-called "summary parameters" that can be measured directly from the DSC-MRI data [1, 5]. Quantification of CBF, CBV, MTT, and Tmax is provided using the perfusion model, which considers time-concentration curves as a convolution of the response function with the arterial input function (AIF). Traditionally, the global AIF is a function that describes the concentration of a contrast agent over time in a brain-feeding artery, such as the middle cerebral artery (MCA) or the internal carotid artery (ICA). However, to overcome the potential suffering from the underestimation of CBF due to delay and dispersion effects, quantification of perfusion parameters can be done by alternatively identifying multiple local arterial regions [6].

Wrong detection of the AIF influences the result of the deconvolution operation and leads to misdiagnosis. Thus, the detection of the AIF is one of the key steps in the quantification of perfusion parameters.

In this work, the existing methods of AIF detection are considered, as well as an assessment of their pros and cons.

**Manual, semi-automatic, and automatic procedures of AIF detection**

There are three main approaches to perform the procedure of AIF detection: manual selection, semi-automatic and automatic detection.

In the case of manual selection of the AIF, the operator should select some number of pixels with a cursor to mark the position of the AIF on the proper imaged slice [6]. To do that, the operator first analyzes a region of interest associated with a brain-feeding artery, such as MCA or ICA. As the AIF should be estimated from inside an artery, the operator interactively examines pixel by pixel various locations in the region of interest. Changes in time-density curves (or time-concentration curves for some software) are displayed as the cursor moves. Based on the shape of the displayed curves the operator marks several suitable pixels deemed to be most arterial-like. It assumes that pixels, which contain primarily arterial contribution, relate to the curves with a large amplitude, small width, and early time-of-arrival [6]. It should be mentioned that the AIF obtained from a single pixel is not reliable enough because of the high sensitivity of the deconvolution method to noise and motion artifacts. Therefore, it is more appropriate to determine AIF as a region or volume [7].

Commonly, the AIF marked according to the above procedure is used as a global input function for tissues in the whole brain. In some cases, to reduce the influence of bolus delay, software for DSC-MRI data analysis provides a possibility to mark arterial pixels in each imaged slice; this means that the operator has to repeat the AIF selection procedure a few times.

Manual selection of the AIF is considered a gold standard. However, performed by a radiologist manual AIF selection is time-consuming by viewing both spatial and temporal DSC-MRI data. Moreover, the measured concentration of contrast agent may vary broadly according to the utilized protocol of contrast agent injection or cardiac output & vascular status of each particular patient. Therefore, manual AIF selection highly depends on the experience and subjective judgments of the operator. It results in not only a highly precise selection of the AIF but also an unrepeatability of measurements and poor consistency in perfusion parameters between studies.

As was mentioned above, the selection of the AIF from a large brain-feeding artery can result in perfusion measurement errors related to delay and dispersion during the bolus transition from the artery to tissues. To overcome mentioned delay and dispersion effects, a local AIF can be selected as close to the tissue of interest as possible [6, 8]. In such a case, a relatively low spatial resolution of DSC-MRI data makes it difficult to identify pixels as representative for small vessels. Besides the difficult selection procedure, the partial volume effect occurs more easily and can cause inaccurate measurements in the case of local AIF selection.

Partially or fully automatic approaches can improve reliability and quality as well as reduce the time needed for the AIF detection procedure.

In the case of a semi-automatic procedure, the detection of the AIF is assisted by highlighting the most representative arterial pixels [6, 9]. Before the operator makes a final determination of the AIF, the software can provide automatic identification of pixels that relate to the curves with the largest amplitude and/or earliest time of arrival. The procedure of AIF detection is also partially automatic when the identification of most arterial-like pixels is provided in a search region that is manually defined by the operator.

The major advantage of a fully automatic procedure of AIF detection is the increasing speed of image processing and, as a result of this, perfusion analysis is performed much faster [6, 7]. It is very important for diagnosing acute stroke. Automatic AIF detection is often performed based on the shape of time-concentration curves [7, 9]. As DSC-MRI data has a relatively low spatial resolution, the partial volume effect leads to shape changes in AIF that can cause errors in operating AIF selection criteria.

Applying software that can perform AIF detection without operator control is a process of 'black box' usage, because extremely difficult to get any quality control on the output of automatic AIF detection. Sometimes the validation of AIF detection can be even more challenging as the searching process is integrated into a more comprehensive automatic system for DSC-MRI data analysis [6].

**Problems of current methods in AIF detection**

To improve reproducibility and make the procedure of AIF detection less user-dependent and much faster, several semi- and fully automatic methods have been proposed.

Mouridsen et al [10] proposed to select automatically the arterial input function using cluster analysis, but since k-means clustering techniques were used, this method showed poor reproducibility.

Yin et al [11] proposed the agglomerative hierarchy clustering algorithm that showed higher reproducibility compare to the k-means clustering. However, further evaluation [12] of Yin's algorithm showed issues associated with detecting AIF within tumorous regions instead of arteries and selecting some truncated or noisy curves as AIF.

To improve the agglomerative hierarchy clustering algorithm, Rahimzadeh et al [12] proposed to exclude non-arterial curves such as tumor, tissue, noisy, and partial-volume affected curves in the pre-processing step. As the authors mentioned, their study population was limited to brain tumor patients. To be applicable for clinical use, the framework with excluding non-arterial curves needs generalization on more diverse diseases such as acute stroke, arterial stenosis, and other brain disorders.

Tabbara et al [13] proposed a multi-stage automatic detection of the local AIF in perfusion MRI, and earlier, they conducted research based on cluster analysis and priority flooding [14]. Despite the good results for both local and global AIF detection that were mentioned in the original research, it was shown the impossibility of applying the proposed method for time series with low temporal resolution [15].

Sobhan [16] proposed to improve the AIF detection by the replacing raw-data-based clustering step of Tabbara's method with feature-based clustering. However, the proposed method targeted global AIF detection and was applicable to only nine glioma patients who did not have any history of arterial abnormalities. Moreover, the method requires some manual operations to create a collection of nominal AIFs in the final step.

One more AIF detection method was developed by Bal et al [9] based on geodesic model of contour-based segmentation that uses a discrete total variation function as geodesic distance to separate homogeneous regions or intensity jumps more accurately. After locating the arterial region, matrix analysis is used to find the AIF as a pixel with maximum peak height within the region. The method showed better arterial features of higher peak position and fast attenuation as compared to the other state-of-the-art methods. However, evaluation of the method based on a contour geodesic model was carried out only on 15 ischemic stroke patients that underwent the same perfusion imaging protocol. Moreover, the proposed method requires some manual operations to set a marker point or to localize the arterial region.

Lipiński and Kalicka [17] proposed a fully automatic process of AIF detection. The method chooses candidates for AIF based on the perfusion descriptors that combine physiological requirements for AIF with mathematical criteria. The final choice of the AIF is made from a class of candidates with the best fit of regression function to the measurements. The efficiency of the proposed method was proved using DSC-MRI measurements from only one 40-year-old male patient, which may lack generalization for clinical applications. One more problem can be outlined: there are variations in the calculation algorithms of characteristics that are used as quality descriptors in the proposed method [5, 18]. Results of different calculation algorithms can vary in terms of accuracy and reliability of DSC-MRI perfusion characteristics.

Fan et al [19] proposed a multi-stream 3D CNN that finds the region used to calculate the average curve as AIF. The network performs feature extracting in both spatial and temporal streams simultaneously. Each stream uses a series of 3D convolutions and the fusion of streams is provided with the linear support vector machine. The model was trained using 100 DSC-MRI in-house collected cases in which 30 cases were healthy and 70 were cases of stroke; AIF annotations were made in the scope of the conducted research.

More recently, Winder et al [20] also proposed a CNN-based pipeline for the automatic AIF selection. The network identifies voxels that represent arterial shapes and geometric averaging is used to get a smooth AIF from multiple selections for one perfusion dataset. The model for MRI-DSC modality was trained using 100 acquisitions from the multicenter stroke imaging study I-KNOW [21]; AIF annotations were made in the scope of the conducted research.

Despite the effectiveness and robustness to detect AIF under conducted tests in both mentioned studies, the accuracy of the proposed CNN-based methods is limited due to 1) the test population was limited to healthy and stroke patients and 2) the process of neural network models obtaining requires the annotation of ground truth as an input [9, 22]. Technically, the last one would mean that annotation mainly would be a manual operation. It might be time-consuming since the operator needs to inspect many pixels to be the most representative arterial input function and to repeat this process for a big number of cases. Manual annotation leads to an inability to guarantee the selection of AIF with optimal characteristics. It also should be mentioned that the datasets used in these studies were private data collections, so there is no possibility to verify and validate the published results.

**Conclusions**

Reliable and accurate perfusion quantification is highly dependent on the AIF, so a proper selection of the AIF is a crucial step for DSC-MRI data analysis.

Manual selection of the AIF is time-consuming and depends entirely on the experience of the operator who is performing it. Moreover, to provide reproducible and robust results, the AIF detection should be done without determining slice, region, or any other actions that are performed by the operator, at the same time, the algorithmic implementation itself should utilize no random initializers in any step of the detection process.

From the other perspective, a fully automatic procedure of AIF detection makes a selection process hidden from the operator. Therefore, software with fully automatic analysis of DSC-MRI data should provide mechanisms to perform quality control of the program output at each step of the analysis.

**References**

1. Calamante, F. (2012). Perfusion magnetic resonance imaging quantification in the brain. In *Visualization Techniques* (pp. 283-312). Humana Press, Totowa, NJ.

2. Boxerman, J. L., Quarles, C. C., Hu, L. S., Erickson, B. J., Gerstner, E. R., Smits, M., ... & Jumpstarting Brain Tumor Drug Development Coalition Imaging Standardization Steering Committee. (2020). Consensus recommendations for a dynamic susceptibility contrast MRI protocol for use in high-grade gliomas. *Neuro-oncology*, 22(9), 1262-1275.

3. Giannatempo, G. M., Scarabino, T., Popolizio, T., Parracino, T., Serricchio, E., & Simeone, A. (2017). 3.0 T perfusion MRI dynamic susceptibility contrast and dynamic contrast-enhanced techniques. In *High Field Brain MRI* (pp. 113-131). Springer, Cham.

4. Digernes, I., Nilsen, L. B., Grøvik, E., Bjørnerud, A., Løvland, G., Vik-Mo, E., ... & Emblem, K. E. (2020). Noise dependency in vascular parameters from combined gradient-echo and spin-echo DSC MRI. *Physics in Medicine & Biology*, 65(22), 225020.

5. Alkhimova, S. M. (2015). Calculation accuracy evaluation of quantitative parameters of overall perfusion assessment. *Eastern-European Journal of Enterprise Technologies*, 6 (9 (78)), 4-9. doi: 10.15587/1729-4061.2015.55908

6. Calamante, F. (2013). Arterial input function in perfusion MRI: a comprehensive review. *Progress in nuclear magnetic resonance spectroscopy*, *74*, 1-32.

7. Bleeker, E. J., van Osch, M. J., Connelly, A., van Buchem, M. A., Webb, A. G., & Calamante, F. (2011). New criterion to aid manual and automatic selection of the arterial input function in dynamic susceptibility contrast MRI. *Magnetic resonance in medicine*, *65*(2), 448-456.

8. Willats, L., Christensen, S., K Ma, H., A Donnan, G., Connelly, A., & Calamante, F. (2011). Validating a local Arterial Input Function method for improved perfusion quantification in stroke. *Journal of Cerebral Blood Flow & Metabolism*, *31*(11), 2189-2198.

9. Bal, S. S., Chen, K., Yang, F. P. G., & Peng, G. S. (2022). Arterial input function segmentation based on a contour geodesic model for tissue at risk identification in ischemic stroke. *Medical Physics*, *49*(4), 2475-2485.

10. Mouridsen, K., Christensen, S., Gyldensted, L., & Østergaard, L. (2006). Automatic selection of arterial input function using cluster analysis. *Magnetic Resonance in Medicine: An Official Journal of the International Society for Magnetic Resonance in Medicine*, *55*(3), 524-531.

11. Yin, J., Yang, J., & Guo, Q. (2014). Evaluating the feasibility of an agglomerative hierarchy clustering algorithm for the automatic detection of the arterial input function using DSC-MRI. *PloS one*, *9*(6), e100308.

12. Rahimzadeh, H., Kazerooni, A. F., Deevband, M. R., & Rad, H. S. (2019). An efficient framework for accurate arterial input selection in DSC-MRI of glioma brain tumors. *Journal of Biomedical Physics & Engineering*, *9*(1), 69.

13. Tabbara, R., Connelly, A., & Calamante, F. (2018). Automatic selection of local arterial input functions in perfusion MRI using cluster analysis and priority-flooding. In *Proc Intl Soc Magn Reson Med* (Vol. 26, p. 2179).

14. Tabbara, R., Connelly, A., & Calamante, F. (2020). Multi-stage automated local arterial input function selection in perfusion MRI. *Magnetic Resonance Materials in Physics, Biology and Medicine*, *33*(3), 357-365.

15. King, A. (2021). *Quantitative Perfusion Measurements in a Novel Large Animal Stroke Model* (Doctoral dissertation, Open Access Te Herenga Waka-Victoria University of Wellington).

16. Sobhan, R. (2020). *Methods for assisting the automation of Dynamic Susceptibility Contrast Magnetic Resonance Imaging Analysis* (Doctoral dissertation, University of East Anglia).

17. Lipiński, S., & Kalicka, R. (2018). Automatic selection of arterial input function in DSC-MRI measurements for calculation of brain perfusion parameters using parametric modelling. *Mathematical Modelling of Natural Phenomena*, *13*(6), 58.

18. Alkhimova, S. Impact of Perfusion ROI Detection to the Quality of CBV Perfusion Map. *Technology Audit and Production Reserves*, *5*(2), 27-30. doi: 10.15587/2312-8372.2019.182789

19. Fan, S., Bian, Y., Wang, E., Kang, Y., Wang, D. J., Yang, Q., & Ji, X. (2019). An automatic estimation of arterial input function based on multi-stream 3D CNN. *Frontiers in neuroinformatics*, *13*, 49.

20. Winder, A., d'Esterre, C. D., Menon, B. K., Fiehler, J., & Forkert, N. D. (2020). Automatic arterial input function selection in CT and MR perfusion datasets using deep convolutional neural networks. *Medical Physics*, *47*(9), 4199-4211.

21. Cheng, B., Forkert, N. D., Zavaglia, M., Hilgetag, C. C., Golsari, A., Siemonsen, S., ... & Thomalla, G. (2014). Influence of stroke infarct location on functional outcome measured by the modified rankin scale. *Stroke*, *45*(6), 1695-1702.

22. de la Rosa, E., Sima, D. M., Menze, B., Kirschke, J. S., & Robben, D. (2021). AIFNet: Automatic vascular function estimation for perfusion analysis using deep learning. *Medical Image Analysis*, *74*, 102211.